\documentclass[12pt]{article}

\input{def.tex}
\newcommand{\nud}{\nu} 

\begin{document}

\title{Fiducial Inference and Decision Theory}


\author{G. Taraldsen\orcidlink{0000-0003-4980-7019}
  and B.H. Lindquist\\
  Norwegian University of Science and Technology, Trondheim, Norway
}

\maketitle

\begin{abstract}
The majority of the statisticians 
concluded many decades ago that
fiducial inference was nonsensical to them.
\citet{HannigIyerLaiEtAl16GeneralizedFiducialInference} and others have,
however,
contributed to a renewed interest and focus.
Fiducial inference is similar to Bayesian analysis, but without requiring a prior.
The prior information is replaced by assuming a particular data generating equation.
\citet{BERGER} explains that Bayesian analysis and statistical decision theory are in harmony.
\citet{TaraldsenLindqvist13FiducialTheoryOptimal} show that fiducial theory and
statistical decision theory also play well together.
The purpose of this text is to explain and exemplify this
together with recent mathematical results.
\footnote{\href{https://zbmath.org/static/msc2020.pdf}{2020 Mathematics Subject Classification}:
  {62-01 Introductory exposition pertaining to statistics;
    62F10 Point estimation;
  62B05 Sufficient statistics and fields;
  62C05 General considerations in statistical decision theory;
58K70 Symmetries, equivariance on manifolds;  
62E10 Characterization and structure theory of statistical distributions;}}
\end{abstract}

\tableofcontents

\newpage

\section{Introduction}\label{sIntroduction}

The fiducial density $\pi (\rho \st r)$ of the correlation $\rho$ given
the empirical correlation $r$ is
\be{dRho}
\pi (\rho \st r) =
\frac{(1 - r^2)^{\frac{\nud - 1}{2}} \cdot
(1 - \rho^2)^{\frac{\nud - 2}{2}} \cdot
(1 - r \rho )^{\frac{1-2\nud}{2}}}
{\sqrt{2} B(\nud + \half, \half)} \cdot
F(\frac{3}{2},-\half; \nud + \half; \frac{1 + r \rho}{2})
\ee
where $B$ is the Beta function, $F$ is the Gaussian hypergeometric function,
and  $n = \nu - 1$ is the sample size \citep{Taraldsen21ConfidenceDensityCorrelation}.
The fiducial is also given by Rao's formula
%
\be{dRhoRao}
\pi (\rho \st r) =
\frac{  (1 - r^2)^{\frac{\nud - 1}{2}} \cdot
  (1 - \rho^2)^{\frac{\nud - 2}{2}} 
}{\pi (\nud - 2)!}
\partial_{\rho r}^{\nud - 2}
\left\{
\frac{\theta - \frac{1}{2}\sin 2\theta}{\sin^3 \theta}
\right\}
\ee
where $\cos \theta = -\rho r$ and $0 < \theta < \pi$.
Rao's formula~(\ref{eqdRhoRao}) is stated as the very last formula in the classical book
{\it Statistical Methods and Scientific Inference}
by \citet[eq.(234)]{FISHER}.
The fiducial for the correlation is also the very first example
of a fiducial distribution presented by \citet{Fisher30InverseProbability}.

It is no longer possible to ask Fisher in person,
but from his writings it seems he would still claim:
\begin{quote}
{\bf  
The fiducial is the answer! }
\end{quote}
We agree.
The problem Fisher gave us is then to explain and characterize relevant questions.
This is a very open-ended problem, and it still needs more investigation.
The original claim of the
naturalness of the fiducial by \citet{Fisher30InverseProbability},
and the many questions and more specific claims that
followed have inspired the development of classical and Bayesian
practical, theoretical, and mathematical statistics.
It still does.

Recent mathematical results that give links between
different modes of inference are presented in an Appendix here without any particular interpretation.
Section~\ref{sDiscussion} includes a brief discussion of related work, and comments on open problems.
The majority of the text is devoted to exemplifying 
the interplay between fiducial arguments and decision theory as
described by \citet{TaraldsenLindqvist13FiducialTheoryOptimal}.
Section~\ref{sExamples} contains examples and Section~\ref{sDecision} presents
decision theory illustrated with a simple example.
In the rest of this Introduction
a direct interpretation of the fiducial distribution is explained.

Consider the following two equivalent equations
\be{RhoR}
r = r(\rho, u) \iff
\rho = \rho(r, u) 
\ee
The first equation gives the uncertainty of the empirical correlation $r$ for a
known correlation $\rho$ from the uncertainty in a Monte Carlo variable $u$.
It is called a data generating equation since computer generated samples from $u$'s
distribution gives samples from the distribution of $r$ when $\rho$ is known.
A data generating equation $r = r(\rho, u)$ follows from a data generating equation for
the binormal.
It is not unique, but $\pi (\rho \st r)$ is unique.

The second equality in equation~(\ref{eqRhoR}) is equivalent with the first equality.
It gives the uncertainty of the correlation $\rho$ for a
known empirical correlation $r$ from the uncertainty in $u$.
It is called a parameter generating equation since computer generated samples from $u$'s
distribution gives samples from the distribution of $\rho$ when $r$ is known.
The corresponding density of $\rho$ when $r$ is known
is given by the two alternative expressions in equation~(\ref{eqdRho}) and
equation~(\ref{eqdRhoRao}) respectively. 
The fiducial distribution is hence simply the result of propagation of the uncertainty as
defined by the uncertainty in $u$ and the generating equations~(\ref{eqRhoR}).

The interpretation of the fiducial in a pure fiducial perspective is as
the state of knowledge of the parameter $\rho$
given the data and the parameter generating model.
This determines then also the state of knowledge of any derived parameter
$\lambda = \lambda (\rho)$.
This interpretation is exactly as for a Bayesian prior and for a Bayesian posterior,
but it is to be used for cases without a prior.
\citet[p.523-4]{TaraldsenLindqvist19DiscussionNonparametricGeneralized}
explain and discuss this point in more detail. 
It is also this interpretation which is stated directly by \citet[p.54-5]{FISHER}:
\begin{quote}
{\em By  contrast, the fiducial argument uses
the observations only to change  the  logical status of
the parameter from one in which nothing is known
of  it, and no probability statement about  it  can
be made,  to  the status of  a  random variable having
a  well-defined distribution.}
\end{quote}

Consider an example where
the result of an experiment is given
by four points with $(x,y)$ coordinates
$(773, 727)$,
$(777, 735)$,
$(284, 286)$, and
$(519, 573)$ \citep{Taraldsen21ConfidenceDensityCorrelation}.
There are reasons a priori for assuming a linear relationship.
This is further supported by 
Figure~\ref{fig1}, and a high value for the
coefficient of determination $R^2 = 97.00 \%$.
\begin{figure}
    \centering
    \includegraphics[width=.8\textwidth]{fig1.pdf}
    \caption{A regression line corresponding
      to an example by \citet[p.434]{Fisher30InverseProbability}.
      } 
    \label{fig1}
\end{figure}
The $R^2$ equals the square of the empirical correlation
$r = 98.49 \%$.
This is an example of linear regression used extensively in applied sciences.

A natural question is:
What about uncertainty?
In fiducial inference the uncertainty is given by the fiducial distribution.
The state of knowledge regarding the correlation is given by the fiducial density shown in Figure~\ref{fig2}.
\begin{figure}
    \centering
    \includegraphics[width=.8\textwidth]{fig2.pdf}
    \caption{The fiducial density for correlation in the \citet[p.434]{Fisher30InverseProbability} example.
      } 
    \label{fig2}
\end{figure}
The uncertainty of any derived parameter $\lambda(\rho)$ follows by propagation of uncertainty.

What about decision theory?
In this case a natural question would be to find a point or interval esimate $\hat{\rho}$ for $\rho$.
Decision theory gives one possible approach.
A possible loss function for a point estimate is the absolute loss
\be{lAbs}
l = \abs{\hat{\rho} - \rho}
\ee
The fiducial risk
\be{rAbs}
\int l(\hat{\rho}, \rho) \pi(\rho \st r) \, d\rho
= \int \abs{\hat{\rho} - \rho} \pi(\rho \st r) \, d\rho
\ee
is minimized by the median $\hat{\rho} = 97.48 \%$.
This is different from the ordinary estimate given directly by $r = 98.49 \%$.

The median is the optimal estimate given the particular loss in equation~(\ref{eqlAbs}). 
In general, the loss depends on the particular application of the analysis.
The absolute loss has the advantage of giving an estimate which is
invariant under one-one transformations of the parameter.

\section{Fiducial decision theory}\label{sDecision}

Given data $y$ it is required to find an {\it action}
$x = x (y)$ that is optimal.
Optimality can be formalized starting with the specification of a {\it loss}
\be{Loss}
l = l (\gamma, x)
\ee
corresponding to a parameter $\gamma$ and the action $x$.
An example for a real statistic is given by squared error loss
\be{LossGS}
l = (\gamma - x)^2
\ee
The estimate $x$ can be a test, a set estimate, a distribution estimate,
or more complicated objects,
and then other loss functions are natural.

It will be assumed in this section that a fiducial distribution
for $\gamma$ is given by the distribution of a random quantity
$\Gamma^y$.
An example was given in the Introduction and further examples will be given.
A more general class of fiducial distributions are given
by Definition~\ref{d2New} in the Appendix.
This class of fiducial distributions includes all the examples in the following.
The random quantity
$\Gamma^y = \gamma (U^a, y)$ is
obtained by solving a data generating equation.
The distribution of $U^a$ is the updated state of
knowledge of $u$ given the data $y$.
The value $a = a(y)$ is the observed value of a maximal invariant statistic $A = a(Y)$.

The distribution of the loss $l (\Gamma^y, x(y))$ for each $y$ determines
if an action $x$ is good.
More generally, the joint distribution of
$l (\Gamma^y, x_A (y))$ and $l (\Gamma^y, x_B (y))$
for each $y$ can be used in a comparison of two actions.
The analysis can sometimes be simplified
by only considering a simpler statistic
given by a single number.
The {\it fiducial risk}, if it exists, is defined to be the expected loss
\be{FidRisk}
r = \E l(\Gamma^y, x)
\ee
The fiducial risk of an action $x = x(y)$ is a statistic $r = r(y)$.
An action $x^*$ is a best fiducial action
if it's fiducial risk $r^*$ is smaller than or equal to any other risk.

A best fiducial action $x = x(y)$ can in good cases be obtained directly
by minimizing the risk.
As a solution of an optimization problem a best fiducial action is also
called an optimal fiducial action.
The prototypical example is given by
\be{OptX}
x(y) = \E \Gamma^y
\ee
which minimizes the risk from the squared error loss in equation~(\ref{eqLossGS}). 
More generally, the optimization problem is much simplified both practically and theoretically
if $x \mapsto l (\gamma, x)$ is convex for all $\gamma$.

An example is given by the maximum likelihood
estimator $y$ of the scale $\theta > 0$ from a random sample of
size $n$ from a gamma distribution with known shape $\alpha > 0$.
The maximum likelihood estimator $y$ of $\theta$ equals the empirical mean of the sample and
the law of $u$ is a gamma distribution with shape
$n \alpha$ and scale $(n \alpha)^{-1}$.
The model generating equation is $y = \theta u$ is given simply by multiplication.
\citet{TaraldsenLindqvist15FiducialPosteriorSampling} investigate this
and some related models in more detail.

Fiducial inference is in this simple case given by solving
$y = \theta u$ with the result
\be{FidSimple}
\theta = y u^{-1}
\ee
The fiducial distribution is hence an inverse gamma distribution scaled by the data $y$.
Exact simulation from this fiducial,
and also the fiducial for any focus parameter $\gamma = \gamma (\theta)$,
is obtained directly from the generator $u$.

In this case the maximal invariant is trivial (a constant), and the
state of knowledge of $u$ is unchanged by the observation.
Formally, in this case, $U^a \sim U$. 
The three examples in Section~\ref{sExamples} demonstrates the more complicated case
where the state of knowledge of $u$ is updated by the data
and $U^a \not\sim U$.

For the loss in equation~(\ref{eqLossGS}) for the gamma scale model
the fiducial optimal action is
\be{FidMeanOpt}
x = \E \Theta^y = \left(\frac{n \alpha}{n \alpha - 1} \right) y 
\ee
assuming $n \alpha > 1$.
It can be observed that this estimate is different from the MLE $y$, but
for large $n \alpha$ the difference is small.

Existence and calculation of a fiducial optimal action is hence similar
to the problem of finding an optimal Bayesian posterior action.
\citet{BERGER} and \citet{Eaton89GroupInvarianceApplications} provide many examples
and theory related to calculating optimal Bayesian actions.
In addition to estimation this includes hypothesis testing and randomized actions.

The next aim is, as in textbooks on Bayesian decision theory, to provide a link
to frequentist decision theory.
The distribution of $l (\gamma, X)$ determines if the action $X = x(Y)$ is good.
A data generating equation can be used to investigate this for each model $\theta$.
This gives a method for comparison of two actions $X_A$ and $X_B$.
Instead of a comparison of the distribution of $X_A$ and $X_B$ one should instead
compare the distribution of the losses $l (\gamma, X_A)$ and $l (\gamma, X_B)$.
It can be observed that a data generating model allows simulation from the
joint law of the actions, and also from the joint law of the losses.

As for the fiducial risk it is convenient, if possible,
to simplify the analysis down to a single number.
The {\it risk} is defined as the expected loss
\be{FidRisk}
\rho = 
\E l(\gamma, X)
\ee
The risk is a function of the model $\theta$
since $\gamma = \gamma (\theta)$ and the distribution of $X$ depends on $\theta$.
The risk of a statistic $X$ is hence a parameter.
The action $X^*$ is {\it uniformly optimal} if the risk $\rho^*$ is 
smaller than or equal to the risk $\rho$ of any other action.
The statement $\rho^* \le \rho$ means that
$\rho^* (\theta) \le \rho (\theta)$ for all models $\theta$.
Existence and uniqueness of a uniformly optimal action is challenging 
since the unknown is the function $x (y)$,
and the bound is to hold uniformly for all models $\theta$.

A much simpler problem is to find 
a fiducial optimal action as defined by minimizing
the fiducial risk.
A candidate action is hence given by a fiducial optimal action.
This turns out to be also uniformly optimal given some additional restrictions
as proved by \citet{TaraldsenLindqvist13FiducialTheoryOptimal}.
The estimators are required to be equivariant and the loss is
assumed to be invariant.
A precise statement is given in Theorem~\ref{tFidOpt} in the Appendix.

The MLE $y$ and the optimal $x$ in equation~(\ref{eqFidMeanOpt}) for the
gamma scale model are both equivariant estimates
in the sense that $x(a y) = a x(y)$.
The loss in equation~(\ref{eqLossGS}) is not invariant, but the losses
\be{InvLoss}
l = \theta^{-2} (\theta - x)^2 
\ee
\be{LogLoss}
l = \left[\ln (\theta) - \ln(x) \right]^2 
\ee
are both invariant for the gamma scale model.
These losses obey $l(a \theta, a x) = l(\theta, x)$.

Uniformly optimal actions corresponding to the losses in
equation~(\ref{eqInvLoss}) and
equation~(\ref{eqLogLoss}) are respectively
given by equation~(\ref{eqFidMeanOpt})
and
\be{ALogLossA}
x = \exp(\E \ln \Theta^y) = \frac{y}{\alpha} \exp \left[\ln (n \alpha) - \psi(n \alpha) \right]
\ee
where $\psi$ is the digamma function.
The action defined by equation~(\ref{eqFidMeanOpt})
is also the equivariant estimate that uniformly minimizes the squared error risk. 
The result in equation~(\ref{eqALogLossA}) is derived by
\citet[p.3759]{TaraldsenLindqvist15FiducialPosteriorSampling}.
It gives a natural estimator for the problem
since the loss equals the squared Fisher information distance
\citep[p.333]{TaraldsenLindqvist13FiducialTheoryOptimal}.

\section{Examples}\label{sExamples}



\subsection{The original fiducial}\label{ssOriginal}

\noindent \citet{Fisher30InverseProbability} introduced
the concept of a fiducial distribution.
Fisher's first example is the fiducial density
$\pi (\rho \st r)$
for the correlation $\rho$ of the binormal distribution.
It is
given by
\be{FFid}
\pi (\rho \st r) = -\partial_\rho F (r \st \rho),
\ee
where $F (r \st \rho)$ is the cumulative distribution function for the
empirical correlation $r$ of a random sample of size $n$ from the binormal distribution.

\citet{Taraldsen21ConfidenceDensityCorrelation} reconsidered this problem
and derived an exact explicit formula for the fiducial density $\pi (\rho \st r)$
as stated in equation~(\ref{eqdRho}).
This theory is here reconsidered and generalized.
The purpose is to demonstrate fiducial inference for a nontrivial problem.
It should be noted that the likelihood is not used in the arguments.
The argument is valid also for models that are not defined by a likelihood.

\begin{example}[Random points in the plane]
  \label{exLower}
Assume that the joint law of
$(u_1,v_1, \ldots, u_n, v_n)$ is known.
A data generating model for $n$ random points in the plane is defined by
\be{Fraser}
  \begin{bmatrix}
    x_i \\
    y_i
  \end{bmatrix}
  =
    \begin{bmatrix}
    \mu_x \\
    \mu_y
  \end{bmatrix}
  +
      \begin{bmatrix}
    \sigma_x & 0 \\
    \rho \sigma_y & \sqrt{1 - \rho^2} \sigma_y
  \end{bmatrix}
\cdot
      \begin{bmatrix}
    u_i \\
    v_i
  \end{bmatrix}
\ee
and $\theta = (\mu_x, \mu_y, \sigma_x, \sigma_y, \rho) \in \Omega_\Theta \subset
\RealN^2 \times [0,\infty)^2 \times (-1,1)$.
\end{example}
A general definition of a data generating equation is
given in  Definition~\ref{d2} in the Appendix.
The data generating model in equation~(\ref{eqFraser}) is equivalent
with generation of $x$-coordinates by $x_i = \mu_x + \sigma_x u_i$ followed
by generation of $y$-coordinates by the linear regression
$y_i = a x_i + b + \sigma v_i$.
Conventional statistical inference takes the resulting probability distribution
for the data as a starting point for analysis.
Fiducial inference includes the given data generating equation
as a part of the model formulation.
This has consequences in the analysis, and for the resulting fiducial.

The slope $a = \rho \sigma_y/\sigma_x$,
the intercept $b = \mu_y - \rho \mu_x \sigma_y / \sigma_x$,
and the
conditional standard deviation $\sigma = \sqrt{1 - \rho^2} \sigma_y$ give
the alternative parameterization $(\mu_x, \sigma_x, a, b, \sigma)$.
The original parameterization
$(\mu_x, \mu_y, \sigma_x, \sigma_y, \rho)$
corresponds to location, scale, and correlation $\rho$
if it is assumed that $\E (U_i V_i) = \E (U_i) = \E (V_i) = 0$ and
$\Var (U_i) = \Var (V_i) = 1$.

Different Monte Carlo distributions and different choices for $\Omega_\Theta$
give different data generating models.
All possibilities in Definition~\ref{d2} can be illustrated.
The case $n=1$ and $\sigma_x, \sigma_y, \rho$ known gives
a conventional, simple, pivotal data generating model.
The case $n > 1$ and $\sigma_x, \sigma_y, \rho$ known can give
a data generating model which is not solvable.
The case $n > 2$ and the largest possible model space can give
a data generating model which is not solvable.
All combinations of known and unknown components for the two alternative
parametrizations give a rich arsenal of different problems for analysis.

A model corresponding to regression of $x$ on $y$ gives an alternative
data generating model.
Fiducial inference for the two models gives different fiducial distributions.
This is in contrast with conventional statistical inference.
Conventional statistical inference gives identical conclusions
for the two models since the probability distribution
for the data is identical if it is assumed that 
$(u_1,u_1, \ldots, u_n, v_n)$ is a random sample from the $\Normal(0,1)$.

If the focus is on the correlation coefficient $\rho$,
then the analysis can be simplified.
The empirical correlation $r$ of
the random points from equation~(\ref{eqFraser}) has a probability
distribution that only depends on $\rho$.
This case was analyzed by \citet[eq.57]{Elfving47SimpleMethodDeducing}
for the binormal case.
Motivated by this we consider a more general case.
\begin{example}[The Elfving Equation]
  \label{exElfving}
  Let the Monte Carlo variabel be a known law for
  $m \in [0, \infty)^2 \times \RealN$,
  let the parameter be the correlation $\rho \in (-1,1)$,
  and let the statistic be the empirical correlation $r \in (-1,1)$.
  The equation
\be{FraserStar}
\sqrt{m_1} \frac{\rho}{\sqrt{1-\rho^2}} -
\sqrt{m_2} \frac{r}{\sqrt{1-r^2}} = m_3
\ee
can be solved to give a unique data generating model
$r = r(m,\rho)$ and a unique parameter generating model
$\rho = \rho(m,r)$.
\end{example}

\begin{theo}[Confidence for correlation]
  \label{t1}
  The fiducial distribution from
  the parameter generating model $\rho = \rho(m,r)$ from equation~(\ref{eqFraserStar})
  is an exact confidence distribution.
  It equals the Bayes posterior for $\rho$ from the model in 
  equation~(\ref{eqFraser}) with either of the priors
\be{Haar}
\pi (\mu_x, \mu_y, \sigma_x, \sigma_y, \rho) = \frac{1}{\sigma_y^2 (1 - \rho^2)}
\ee
and also from the prior density
\be{HaarU}
\pi (\mu_x, \mu_y, \sigma_x, \sigma_y, \rho) = \frac{1}{\sigma_x^2 (1 - \rho^2)}
\ee
where the law of $m$ is determined by the conditional
law of $(u_1,v_1, \ldots, u_n, v_n)$ given the maximal invariant $a$ for the group
action in  equation~(\ref{eqFraser}).
If $(u_1,v_1, \ldots, u_n, v_n)$ is a random sample from $\Normal(0,1)$,
then 
$m_1 \sim \chi^2 (\nud), m_2 \sim \chi^2 (\nud - 1), m_3 \sim \Normal(0,1)$
are independent with $\nud = n-1 > 1$,
and the fiducial density $\pi (\rho \st r)$ is given by equation~(\ref{eqdRho}).
The one-sided confidence intervals from $\pi (\rho \st r)$ are then
uniformly most accurate invariant with respect to the location-scale groups on the two coordinates.
\end{theo}
\begin{proof}
  The last part is proved by \citet{Taraldsen21ConfidenceDensityCorrelation}.
  The first claim follows from Theorem~\ref{t3}.
  Equality of the fiducial and the posterior follows from  Theorem~\ref{t4}
  applied to the conditional model given $a$.
  It is important to note that the fiducial is not only given by $r$, but it depends also
  on the maximal invariant $a$.
  The Basu theorem ensures equality in the normal case since in this case the
  $r$ is a part of a complete sufficient statistic so $r$ and $a$ are independent.
  In this case the fiducial depends only on $r$ as given by $\pi(\rho \st r)$.\\
$\;$ \hfill  \QED  
\end{proof}

\subsection{The scaled uniform model}\label{sConditional}

Consider the data generating equation
\be{dgU}
y = \theta u
\ee
where $\theta > 0$ and
the generator $u = (u_1, u_2)$ is given by the largest and smallest
observation from a random sample of size $n$ from the
uniform distribution on the interval $(1-k,1+k)$.
The corresponding statistical model was recently discussed by
\citet{Mandel20ScaledUniformModel} and before that by
\citet{GaliliMeilijson16ExampleImprovableRao}.
The existence of optimal estimators was left open by their analysis.
The minimal sufficient statistic $y$ is not complete,
but unique optimal equivariant decision rules exist.
This was proved by
\citet{Taraldsen20FiducialSymmetryAction} and summarized by
\citet{Taraldsen20MichaMandel2020}.

The data generating equation does not have
a measurable selection solution for all data $y$ and all generators $u$.
It is an over determined system with two equations for the single unknown parameter.
This is a general problem in fiducial inference,
and many different solutions have been suggested.
One approach is to replace the original data with a minimal sufficient statistic as
for the scaled gamma model.
The problem was there conveniently formulated directly with the minimal sufficient statistic.
This does not always solve the problem as exemplified here with the {\it scaled uniform model}.
The minimal sufficient is two dimensional, but the model space is one dimensional.

A second approach is to expand the parameter space so that solutions exist,
and then condition the fiducial down to the original space.
In the present example a fiducial is obtained naturally for a parameter $(\theta_1, \theta_2)$.
The original model is obtained by restriction to the curve
where $(\theta_1, \theta_2) = (\theta, \theta)$.
A final fiducial can be obtained by conditioning on
$C(\Theta) = \Theta_1 - \Theta_2 = 0$ or by $C(\Theta) = \Theta_1/\Theta_2 = 1$,
or by an infinity of other possibilities.
The result will depend on the choice of the condition $C (\Theta)$,
and the function $C$ is then part of the model assumptions.
This approach is introduced and discussed in more detail by
\citet{TaraldsenLindqvist17FiducialString}
and by \citet{TaraldsenLindqvist18ConditionalFiducialModels}. 
A related approach is used by \citet{CisewskiHannig12GeneralizedFiducialInference}.

A third approach, which is related to the previous,
is to condition on the generator $U$ so that a solution for the parameter exists.
This is the approach explained after Definition~\ref{d2New} and indicated previously.
One possibility is to rely on a particular choice of an
ancillary statistic $A = a (Y)$.
Assume that $a$ is invariant in the
sense of $a (\theta u)$ being a function of only $u$. 
This can be used to replace the original model by a
conditional fiducial model given the condition
\be{CondFid}
a (\theta^* U) = a
\ee
where $a = a(y)$.
The law of $U^a$ defined by this condition is the updated state of knowledge
about $u$ given by the data $y$.
The choice of $\theta^*$ in equation~(\ref{eqCondFid}) does not influence
the fiducial since
$a (\theta U)$ does not depend on $\theta$.
A fiducial is then defined as the distribution of
$\Theta^y = \theta (U^a, y)$.

A natural choice for the ancillary would be to choose a maximal invariant:
Any other invariant is then, by definition, a function of this invariant.
Conditions for existence and uniqueness (up till 1-1 correspondences) of a maximal invariant
is left for future investigations.
\citet{Basu64RecoveryAncillaryInformation}
demonstrates that the choice of an ancillary can be problematic,
but \citet{Cox71ChoiceAlternativeAncillary} gives a possible resolution for some particular cases.
Conditioning on an ancillary to obtain a fiducial
is also discussed briefly by \citet[p.1350]{HannigIyerLaiEtAl16GeneralizedFiducialInference}.
A fully satisfactory solution exists, however, in the
group case, as explained by \citet{TaraldsenLindqvist13FiducialTheoryOptimal}. 

For the scaled uniform model, a unique maximal invariant is given by
\be{MaxA1}
a (U) = U_2/U_1
\ee
The fiducial is given by the distribution of
$U_1 \st (a (U) = y_2/y_1)$.
Calculus gives that it is $\Pareto$
with truncation interval $[(1-k)/a, (1+k)]$
and index $-n$ .
The fiducial is then the corresponding distribution of
$\Theta = y_1 / U_1$.
It is $\Pareto$ with
index $n$ and truncation interval $[\thetaml, \thetamu]$,
so
\be{ParetoPDF}
\Theta^y \sim \pi (\theta \st y) \propto
(\thetaml \le \theta \le \thetamu) \cdot \theta^{-n - 1}
\ee
where $\thetaml = y_1 /(1+k)$ and $\thetamu = y_2 /(1-k)$
\citep{Taraldsen20FiducialSymmetryAction}. 

The conditional model just described fulfills all the assumptions
of Theorem~\ref{tFidOpt}.
Uniformly optimal actions for the conditional model
corresponding to the losses in
equation~(\ref{eqInvLoss}) and
equation~(\ref{eqLogLoss}) are respectively
\be{thetaOpt}
x = \frac{\E ([\Theta^y]^{-1})}{\E ([\Theta^y]^{-2})}
= \frac{\alpha}{\alpha - 1} \cdot
\frac{1 - {b_*}^{1 - \alpha}}{1 - {b_*}^{-\alpha}}
\cdot \thetaml
\ee
\be{OptLog}
x = \exp \E \ln \Theta^y = 
\exp \left\{
  \frac{1}{n} -
  \frac{\ln b_*}{1-{b_*}^n}
\right\} \cdot
\thetaml
\ee
with $b_* = \thetamu/\thetaml$ and $\alpha = n + 2$. 
Both actions are uniformly optimal actions.

The action defined by equation~(\ref{eqthetaOpt}) is
also the action that minimizes the expected squared error
in the class of equivariant estimators.
The action defined by equation~(\ref{eqOptLog}) is,
however, possibly preferable for some investigations since it minimizes also the 
loss corresponding to the squared Fisher information distance
\citep[p.333]{TaraldsenLindqvist13FiducialTheoryOptimal}.


\subsection{Optimal linear prediction}\label{linmod}

Consider a data generating equation
\be{DGLin}
y = \theta + u
\ee
where $y$ and $u$ belong to a Hilbert space,
and $\theta$ is restricted to be in a closed subspace
$\Omega_\Theta$.
A known distribution for $u$ defines then a
conventional fiducial model.
The prototypical example is given by $\theta = X \beta$ where $\beta$ are
unknown regression coefficients, $X$ is the design matrix,
and $\Omega_\Theta$ is the range of $X$.
The case where the dimension of $\beta$ is larger than the dimension
of $y$ has received considerable attention.
The analysis that follows gives an optimal equivariant action for a
prediction problem.

The real or complex Hilbert space may be infinite dimensional
as discussed by \citet[p.331]{TaraldsenLindqvist13FiducialTheoryOptimal} for the case
where $\Omega_\Theta$ equals the Hilbert space.
In this sense this exemplifies a non-parametric fiducial model.
The case with $\Omega_\Theta$ being a strict subspace can be treated
by conditioning on a maximal invariant as explained next.
The analysis here is an extension of the analysis given by
\citet[p.331]{TaraldsenLindqvist13FiducialTheoryOptimal}.
It exemplifies in particular that the fiducial can be non-unique,
but that a unique optimal decision can be derived anyway.

Let $p$ be the orthogonal projection on $\Omega_\Theta$.
It follows that $a = 1 - p$ is a maximal invariant.
This is the projection on the orthogonal complement of $\Omega_\Theta$.
The law of $U^a$ is determined by conditioning $U$ on
\be{U1}
a (U) = a
\ee
where $a = a(y)$ is the observed value of the maximal invariant.
The law of $U^a$ is the updated state of knowledge of $u$
given by the observation.

The fiducial is simply the law of
\be{F1}
\Theta^y = y - U^a
\ee
In the case of i.i.d. Gaussian $U$ it follows that
$U^a = a + p U$,
but this does not hold in general.

Consider the parameter $\gamma = A \theta$ where
$A: \Omega_\Theta \into \Omega_\Gamma$ is a bounded linear operator.
This includes the case $A=I$,
but also the case when $\Omega_\Gamma$ is one dimensional.
The action of $\theta$ on $\Omega_\Gamma$ is given by 
  \be{GroupActionLinear}
    \theta (\gamma) = A \theta + \gamma
  \ee
The loss
  \be{LossLinear}
     l = \abs{\hat{\gamma} - \gamma}^2
  \ee
is invariant.
A fiducial optimal estimate of $\gamma = A \theta$ is
\be{FidDGLin}
\hat{\gamma} = \E \Gamma^y = \E (A \Theta^y)
\ee
Theorem~\ref{tFidOpt} in the Appendix gives 
that this is the uniformly optimal equivariant estimate.
This is an equivariant estimator,
$\gamma (\theta + y) = A \theta + \gamma (y)$, since
  \be{GroupActionEquiv}
\Theta^{\theta + y} = \theta + \Theta^y
  \ee

  Let $x_*$ be given and consider the prediction of
  a real $Y_* = x_* \beta + U_*$.
  Assume that $\E U_* = 0$, and that the distribution of $u$ and $u_*$ are independent.
  Assume furthermore that there exist an $A$ such that
  $A X = x_*$, and if not let $A$ be the least squares solution found by projection.
  The previous argument gives an optimal estimate
  of $\gamma = A X \beta$.
  If $A X = x_*$, then this is an optimal estimate of the
  best linear prediction $x_* \beta$ of $Y_*$.

Many other cases can be considered and treated smoothly along the indicated path.
One case is given by replacing $u$ by $\sigma u$ where $\sigma$
is also unknown.
It should also be noted that the case
$\theta = X \beta$ can give examples of a non-unique
fiducial distribution for $\beta$,
but they all give the same unique optimal estimator $\hat{\gamma}$.
\citet{Fraser79inferenceAndLinear} demonstrates fiducial inference
for many other possible cases.




\section{Discussion}\label{sDiscussion}

Fiducial inference is generally speaking not a well-defined concept.
Different authors give different definitions.
The adjective "fiducial" comes from the Latin {\it fiducia} for faith
and means something taken as standard of reference or
something founded on faith or trust.
The term "fiducial" has been used in the previous also as a noun instead of the longer "fiducial distribution''.
This is in harmony with the tradition of referring to the 
``posterior'' used as a noun instead of using the longer "posterior distribution''.

\citet[p.147]{Pedersen78fiducialDead} concluded
that the fiducial argument has had very limited success and that it was essentially dead.
There was then hence little confidence in fiducial inference among most statisticians.
This is also true today.
Fiducial inference is viewed by many statisticians today as a
somewhat obscure topic.
Different versions are presented in textbooks and reference works,
and most often with critical remarks.
\citet[p.291]{CasellaBerger02StatisticalInference},
\citet[p.134-]{KendallStuart61AdvancedTheoryStatistics},
and
\citet[p.440-]{StuartOrdArnold99KendallAdvancedTheory}
present versions based directly on the likelihood,
but \citet[p.246]{CoxHinkley74TheoreticalStatistics},
\citet[p.77]{Sprott00StatisticalInferenceScience},
and \citet{Barnard95fiducial}
present versions based on pivotal quantities.

Fiducial inference, as presented here, is a generalization of the version based
on pivotal quantities.
It is in particular not restricted to models defined by a likelihood.
It is based on a data generating model.
A data generating model
is a most convenient starting point for
inference since data $y$ can be simulated by simulating $u$ on a computer for
each given model $\theta$.
A data generating model gives also a data generating model for any 
statistic by $x = x(y(u,\theta)) = x(u, \theta)$,
and its properties can be determined by simulation.
Properties of the data generating model $x = x(u,\theta)$ of the statistic $x$
decides if the statistic $x$ is good for inference.
This is one advantage of a data generating model as compared with a statistical model given
by a specification of the conditional law of the data directly.

\citet{Eaton89GroupInvarianceApplications} demonstrates the importance of
symmetry considerations in statistics.
Symmetry is also important for fiducial inference
as discussed in considerable detail by \citet{FRASER}.  
A more complete review of fiducial inference, including it's history,
is beyond the scope of this text, but
\citet{Seidenfeld79PhilosophicalProblemsStatistical},
\citet{DawidWang93FiducialPredictionSemiBayesian},
\citet[p.175]{LehmannRomano05testing},
\citet{VeroneseMelilli15FiducialConfidenceDistributions},
\citet{HannigIyerLaiEtAl16GeneralizedFiducialInference},
\citet{TaraldsenLindqvist18ConditionalFiducialModels},
and \citet{Dawid20FiducialInferenceThen}
can be consulted for discussion and further references.

\citet[p.50]{FRASER} refers to the generator variable $u$ as the
{\it error variable} and the data generating equation as the {\it structural equation},
but then in a more restrictive setting where the model $\theta$ is given by
a 1-1 transformation of the data space $\Omega_Y = \Omega_U$.
The data generating model is then $y = \theta u$ where $\theta$ belongs to a group
and is acting on the data space $\Omega_Y = \Omega_U$.
The notation $y = \theta u$ is also used by \citet{DawidStone82} for
the more general case of a jointly measurable mapping
of the {\it model} parameter $\theta \in \Omega_\Theta$
and the {\it generator} variable $u \in \Omega_U$ to
the {\it data} $y \in \Omega_Y$.

Consider again the model given by equation~(\ref{eqFraser}).
The case with $\mu_x, \mu_y, \sigma_x, \sigma_y$ known is particularly challenging.
It gives a model which is not solvable even after conditioning on a maximal invariant.
The resulting statistical model for the data given by assuming 
that the Monte Carlo variables are a random sample from the standard normal was
used by \citet[p.11]{Basu64RecoveryAncillaryInformation}
to illustrate the difficulty with some of the arguments by \citet{Fisher35LogicInductiveInference}.
The $x$ coordinates is an ancillary statistic,
and the conditional model for the $y$ coordinates can be solved.
Interchanging the roles of the coordinates gives a different solution.
Neither solutions are in fact good from a frequentist point of view.
\citet{Helland-Moe21ObjectiveInferenceCorrelation} considers many alternative solutions.
The best solution is obtained by considering the larger two-parameter model 
$\theta = (\sigma, \rho)$ with $\mu_x = \mu_y = 0$, $\sigma_x = \sigma_y = \sigma$.
The resulting fiducial is in particular better than the fiducial given by equation~(\ref{eqdRho}),
so the information $\sigma_x = \sigma_y$ does give
improved inference for the correlation.
An alternative data generating model was used to get this improvement.
It remains a challenge to obtain 
an improved frequentist solution using the additional information given by $\sigma = 1$.

\appendix

\section{Mathematical Definitions and Theorems}\label{sMath}

The purpose of this section is to briefly summarize mathematical definitions
and theorems providing
links between fiducial inference, Bayesian inference, and confidence distributions.
Any {\it function} used in the following is assumed to be measurable.
Any {\it set} is likewise assumed to be a {\it measurable space} \citep[p.8]{RUDIN}.
The qualifier {\it measurable} is henceforth usually dropped.
A statistic $x = x(y) \in \Omega_X$ is a (measurable!) function of the data $y \in \Omega_Y$.
A particular sample space $\Omega_X$ of a particular statistic
$X$ is sometimes called the action space in the context of decision theory
to differentiate this from all other statistics.
An action is a statistic.

The initial assumption for conventional statistical analysis is a
specification of an indexed family $\{\pr_Y^\theta \st \theta \in \Omega_\Theta\}$ of
probability distributions on the data space $\Omega_Y$.
It is assumed that $Y \sim \pr_Y$,
but the model $\theta \in \Omega_\Theta$ is unknown. 
A parameter $\gamma = \gamma(\theta) \in \Omega_\Gamma$
is a function of the model $\theta$.
Both the parameter space $\Omega_\Gamma$ and the model space $\Omega_\Theta$
are measurable spaces as explained above.
A particular parameter $\gamma$ is sometimes called the focus parameter
to differentiate this from all other parameters.
In any particular problem all of the above spaces with it's structure must be
specified in a mathematical analysis of the problem.

\idx{Fiducial inference},
as formulated mathematically by \citet{TaraldsenLindqvist13FiducialTheoryOptimal,%
  TaraldsenLindqvist15FiducialPosteriorSampling,%
  TaraldsenLindqvist19DiscussionNonparametricGeneralized},
and \citet{Taraldsen21ConfidenceDensityCorrelation},
is inference where a given data generating model is part
of the problem formulation.
Fiducial inference is then different from both Bayesian and frequentist inference
since neither are based on the concept of a data generating model.
\begin{Def}[Generating Models]
  \label{d2}
  A \idx{data generating model} for a statistic $x$ is given by a function
  \be{DGx}
  x = x (\gamma, u)
  \ee
where $\gamma = \gamma (\theta)$ is a parameter.
The law of the \idx{Monte Carlo variable} $u$ 
is given by a family of probability distribution
$\{\pr_U^\theta \st \theta \in \Omega_\Theta \}$
indexed by the \idx{model parameter} $\theta \in \Omega_\Theta$. 
A data generating model is \idx{conventional} if
the law $\pr_U$ of the Monte Carlo variable is known.
The data generating model is \idx{solvable} if it can be solved to give
a \idx{parameter generating model}
\be{PG}
\gamma = \gamma (x, u)
\ee
A data generating model is \idx{simple} if it
is solvable with a unique solution.
The model is \idx{pivotal} if it is conventional and can be uniquely solved to give
a \idx{pivotal generating model}
\be{Ug}
u = u (x, \gamma)
\ee
\end{Def}

It has been explained and exemplified
how a fiducial distribution can be
determined by a data generating model.
It will always be defined as the distribution of a parameter generating model.
The simplest case is given by a solvable model.

\begin{Def}[Fiducial]
  \label{d2New}
Assume a data generating model $x = x(\gamma, u)$ is a solvable and that
$\gamma = \gamma (x, u)$ is a corresponding parameter generating model.
A \idx{fiducial distribution} for $\gamma$ is then defined to be the distribution of
$\gamma (x, U)$ for fixed $x$.
\end{Def}

Definition~\ref{d2New} gives a unique fiducial from any simple conventional data generating model.
Different choices of measurable selection solutions of equation~(\ref{eqDGx})
give different fiducials if the model is only solvable.
If the data generating model is solvable conditionally
given a maximal invariant $a (X) = a$,
then a \idx{fiducial distribution} for $\gamma$ is defined to be the distribution of
$\gamma (x, U^a)$ for fixed $x$.

There are many different data generating models for a given statistical model.
In general, as demonstrated, the corresponding fiducial depends on the
data generating model.
For the case of a real statistic,
exemplified by the empirical correlation $r$,
the fiducial is, however,
uniquely determined by the statistical model under mild assumptions.

\begin{theo}[Uniqueness]
  \label{t3}
  The fiducial distribution from of a real valued, strictly monotonic, and simple
  data generating model is uniquely determined by the sampling distribution of the statistic.
  If, additionally, the sampling distribution of the statistic is continuous,
  then the fiducial distribution is an exact confidence distribution.
\end{theo}
\begin{proof}
  This is Theorem~1 proved by \citet[p.143]{TaraldsenLindqvist18ConditionalFiducialModels}.
  The idea is that inversion of the cumulative distribution function
  gives a data generating model which is both simple and pivotal and a calculation
  shows that the fiducial from this coincides with the initial model.
\QED  
\end{proof}

As explained by \citet[p.329]{TaraldsenLindqvist13FiducialTheoryOptimal}
it is known that the fiducial coincides with
the posterior from a right Haar prior for a group model.
A more general result was proven by \citet[p.3756, Thm 2.1]{TaraldsenLindqvist15FiducialPosteriorSampling}.
\begin{theo}[Bayes]
  \label{t4}
  The fiducial from a simple data generating model $t = t(\theta,u)$
  is the Bayes posterior from a $\sigma$-finite prior $\pr_\Theta$ if the distribution of
  $t (\theta, u)$ does not depend on $u$ when $\theta \sim \pr_\Theta$.
\end{theo}
\begin{proof}
  The idea of the proof by \citet[p.3765]{TaraldsenLindqvist15FiducialPosteriorSampling} is given by
  establishing a joint density for $(t(\Theta, U),U)$ using  
  the Fubini theorem together with the general change-of-variables theorem from measure theory.
  This is different from the more common change-of-variables
  theorem which requires differentiability.
  It should also be noted that there is no assumption of existence of a dominating measure.
  The proof holds in the generality as stated.
  It follows also that the posterior is proper.
\QED  
\end{proof}

The assumption of Theorem~\ref{t4} is fulfilled when
$t = t(\theta, u) = \theta u$ is given by group multiplication
and $\pr_\Theta$ is chosen as the right Haar prior.
The priors in Theorem~\ref{t1} correspond to the right Haar priors
corresponding to the choice of a Cholesky decomposition in lower- or upper-triangular matrices.
Explicit expressions for right Haar priors are derived and discussed
by \citet{Eaton89GroupInvarianceApplications}
in the more general context of group invariance in statistics.
It is noteworthy that the fiducial solution does not need the expressions for the priors.
The fiducial provides hence a simplified approach for Bayesian analysis for this case.

\begin{Def}[Confidence]
  \label{dCD}
  A confidence distribution $\C$ for a parameter $\gamma$ is a
  distribution estimator with 
  \be{CD}
  \C (A_p) = p
  \ee
for a family of confidence sets $A_p$ with level $p$ for all levels $0 < p < 1$.
If equation~(\ref{eqCD}) holds only for all $p \in I \subset ]0,1[$, then
$\C$ is a confidence distribution with level set $I$.
\end{Def}
\begin{theo}[Pivotal]
  \label{t5}
  The fiducial from a conventional simple pivotal data generating model $t = t(\gamma, u)$
  is an exact confidence distribution if $\pr_U$ is continuous. 
\end{theo}
\begin{proof}
  Let $\pr (U \in B) = p$ for an event $A \subset \Omega_U$.
  Define $A (t) = \{\gamma \st u = u(\gamma, t) \in B \}$.
  It follows that $\pr (\gamma (t,U) \in A (t)) = p$ so equation~(\ref{eqCD})
  is fulfilled.
  Furthermore, $\pr (\gamma \in A(T)) = p$, so $A$ is an exact confidence set with level $p$. 
  The set of possible levels $I$ is determined by $\pr_U$, but
  $I =  ]0,1[$ when the Monte Carlo variable is continuous.
\QED  
\end{proof}

\citet[p.334]{TaraldsenLindqvist13FiducialTheoryOptimal} observe that data generating
models can be used to derive confidence distributions.
The assumed existence in Definition~\ref{dCD} of a
family $\{A_p \st p \in I\}$ of confidence sets is important,
but the family is not unique.
Definition~\ref{dCD}
is discussed in more technical detail by \citet{Taraldsen21JointConfidenceDistributions}.
The construction of suitable confidence sets associated with a confidence distribution
is an important problem.
This is discussed by
\citet{LiuLiuXie21NonparametricFusionLearning}
using the concept of a depth-CD.

It is seen from the previous that a fiducial can be a
Bayes posterior and it can also be a confidence distribution.
The fiducial can also determine optimal decision rules.

\begin{theo}
  \label{tFidOpt}
The risk of an equivariant rule in a group invariant problem is
determined by a fiducial distribution if the group acts transitively on the model parameter space.
\end{theo}
\begin{proof}
This is Theorem~1 in \citet{TaraldsenLindqvist13FiducialTheoryOptimal}.
\end{proof}

It should be noted that a fiducial distribution always exists in the previous statement,
but uniqueness of a fiducial distribution is not assumed. 
Optimal inference procedures are determined by the fiducial distribution regardless of
the choice of a measurable selection for the determination of a fiducial distribution.
This is exemplified by the linear prediction example by cases where
the null space of the design matrix is nonzero.


\bibliographystyle{chicago}
\bibliography{bffbook} 


\end{document}